\documentclass{WileyMSP-template}
\pdfoutput=1

\usepackage{graphicx}
\usepackage{acronym}
\usepackage{amsmath,bm}
\usepackage{cleveref}
\usepackage{siunitx}
\usepackage{xspace}

\newacro{ABF}{artificial bacterial flagellum}
\newacroplural{ABF}[ABFs]{artificial bacterial flagella}

\newacro{CMA-ES}{derandomised evolution strategy with covariance matrix adaptation}

\newacro{DPD}{dissipative particle dynamics}

\newacro{ODE}{ordinary differential equation}

\newacro{RL}{reinforcement learning}

\newcommand{\RE}{\mathrm{Re}}

\newcommand{\norm}[1]{\left\lVert#1\right\rVert}

\DeclareMathOperator*{\argmin}{arg\,min}

\newcommand{\sm}{supplementary material\xspace}
\newcommand{\software}[1]{\textit{#1}}

\newcommand{\gkey}[1]{\protect\includegraphics{./figures/keys/#1}\kern-2ex}

\begin{document}

\pagestyle{fancy}
\rhead{\includegraphics[width=2.5cm]{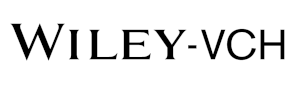}}

\title{Independent Control and  Path Planning of Microswimmers with a Uniform Magnetic Field}

\maketitle

\author{Lucas Amoudruz}
\author{Petros Koumoutsakos*}

\begin{affiliations}
L. Amoudruz, Prof. P. Koumoutsakos\\
Computational Science and Engineering Laboratory, ETH Z\"{u}rich, CH-8092, Switzerland.\\
Email: petros@seas.harvard.edu \\

L. Amoudruz, Prof. P. Koumoutsakos\\
John A. Paulson School of Engineering and Applied Sciences, Harvard University, Cambridge, MA, USA.

\end{affiliations}


\begin{abstract}
  Artificial bacteria flagella (ABFs) are magnetic helical micro-swimmers that can be remotely controlled via a uniform, rotating magnetic field.
  Previous studies have used the heterogeneous response of microswimmers to external magnetic fields for achieving independent control.
  Here we introduce analytical and reinforcement learning control strategies for  path planning to a target by  multiple swimmers using a uniform magnetic field.
  The comparison of the two algorithms shows the superiority of reinforcement learning in achieving minimal travel time to a target.
  The results demonstrate, for the first time, the effective independent navigation of realistic micro-swimmers with a uniform magnetic field in a viscous flow field.
\end{abstract}

\section{Introduction}
\label{se:introduction}

The magnetic control of micro-swimming devices \cite{Cao2014,Yang2020AnnRev,tierno2008,liu2018,liao2019} through micro-manipulation \cite{zhang2010artificial,yu2020}, targeted drug delivery \cite{mhanna2014artificial,sharan2021} or convection-enhanced transport \cite{Schuerle2019}, has created new frontiers for bio-medicine.
A particularly  promising technology involves corkscrew-shaped magnetic micro-swimmers (\acp{ABF}) that propel themselves  when subjected to a rotating magnetic field \cite{zhang2009artificial}.
Rotating magnetic fields can form propulsive gradients and they are arguably preferable to alternatives, such as electric fields, for in-vivo conditions \cite{Gu2019,bente2018}.
However, the independent, yet coordinated, control of individual \acp{ABF} is challenging as it requires balancing between the magnetic forces and the hydrodynamic interactions between the swimmers while the employed magnetic fields are practically uniform over lengths of few micrometers.
We note that independent navigation of mm-sized micro-swimmers has been shown in \cite{Wong2016} through experiments and simulations, while in \cite{Colabrese2017} a \ac{RL} algorithm was applied to adjust the velocity of an idealized swimmer in simulations with one way coupling with a complex flow field.
Control of swimmers using two-way coupling and \ac{RL} have been demonstrated with linked-spheres at low Reynolds numbers \cite{Tsang2020} and for artificial fish in  macro-scales \cite{Verma2018}.
Similarly, genetic algorithms have been used to navigate micro-swimmers towards high concentrations of chemicals \cite{hartl2021}.

The problem of heterogeneous micro-robots navigation via a uniform input has been studied in two dimensions on surfaces \cite{floyd2011} and in a fluid at rest \cite{vach2015steering}.
The steering of two micro-propellers along two distinct paths in 3 dimensions has been accomplished with the help of magnetic fields gradients \cite{diller2013}.
These advances exploited the heterogeneous response of micro-swimmers to a uniform input to achieve independent trajectories along a prescribed path.
These control methods are based on short horizon objectives (stay on the prescribed path) and do not provide the trajectory that minimizes the travel time to a target position, particularly in the presence of a background flow.
In addition, strong background flows restrict the set of feasible paths for given micro-swimmers.
To the best of our knowledge the steering of multiple micron-sized swimmers towards a target in a minimal time under a background flow and a uniform magnetic field has not been reported before.

In this work, we present two methods to independently guide two micro-\acp{ABF} towards a single target in the presence of a uniform magnetic field.
The two methods rely on simulations of swimming \acp{ABF} using an \ac{ODE} model.
The model is calibrated with the method of \ac{DPD} \cite{espanol1995statistical,alexeev2020a}, taking into account the particular geometry of the swimmers and their interactions with the viscous fluid.
We first present a semi-analytical solution for the simple yet instructive setup of multiple, geometrically distinct \acp{ABF} in free space,  with zero background flow.
This result enables understanding of the design constraints for the \acp{ABF} necessary for independent control and how their geometric characteristics relate to their travel time.
We then employ \ac{RL} to control multiple \acp{ABF} trajectories in a broad range of flow conditions including a non-zero background flow.

\section{Artificial bacterial flagella}
\label{se:ABF}

The \acp{ABF} are modeled as microscopic rigid bodies of length $l$ with position $\mathbf{x}$ and orientation $q$ (represented by a quaternion), immersed in a viscous fluid and subjected to a rotating, uniform, magnetic field.
We estimate that the magnetic and hydrodynamic interactions between \acp{ABF} are orders of magnitude smaller than those due to the magnetic field for dilute systems (see \sm) and we ignore inertial effects due to their low Reynolds number ($\RE \approx \num{e-3}$).
Following this approximation, the system is fully described by the position and orientation of the \acp{ABF}.


Additionally, the linear and angular velocities of the \ac{ABF}, $\mathbf{V}^b$ and $\bm{\Omega}^b$, are directly linked to the external force and torque, $\mathbf{F}^b$ and $\mathbf{T}^b$, via the \textit{mobility matrix} \cite{happel1981low},
\begin{equation} \label{eq:mobility}
  \begin{bmatrix}
    \mathbf{V}^b \\
    \bm{\Omega}^b
  \end{bmatrix}
  =
  \begin{bmatrix}
    \Delta & Z \\
    Z^\text{T} & \Gamma
  \end{bmatrix}
  \begin{bmatrix}
    \mathbf{F}^b \\
    \mathbf{T}^b
  \end{bmatrix},
\end{equation}
where the superscript $ ^b$ indicates that the quantity is expressed in the \ac{ABF} frame of reference, for which $\Delta$, $Z$ and $\Gamma$ are diagonal.
The matrices  $\Gamma$ and $Z$ represent the application of torque to changing the angular and linear velocity, respectively.
The \acp{ABF} are propelled by torque applied through a magnetic field and we assume that it can swim only in the direction of its main axis so that $Z$ has only one non-zero entry ($Z_{11})$.
The coefficients in the mobility matrix are often estimated by empirical formulas for low Reynolds number flows \cite{kim2013microhydrodynamics}.
Here we estimate the components of the mobility matrix for the specific ABF by conducting flow simulations using \ac{DPD} \cite{espanol1995statistical,alexeev2020a}, which we validate against experimental data of \cite{mhanna2014artificial} (see \sm).
We remark that the shape (pitch, diameter, length, thickness) of the \ac{ABF} influence the elements of these matrices and the present approach allows to  account for these geometries.

The \ac{ABF} with a magnetic moment $\mathbf{m}$ is subjected to a uniform magnetic field $\mathbf{B}$ and hence experiences a torque
\begin{equation} \label{eq:magn:torque}
  \mathbf{T} = \mathbf{m} \times \mathbf{B}.
\end{equation}
No other external force is applied to the \ac{ABF}, hence $\mathbf{F} = \mathbf{0}$.
Combining \cref{eq:mobility} with the kinematic equations for a rigid body gives the following system of \acp{ODE}:
\begin{subequations} \label{eq:ODE:full}
  \begin{align}
    \dot{\mathbf{x}} &= \mathbf{V}, \label{eq:ODE:x} \\
    \dot{q} &= \frac{1}{2} q \otimes \hat{\bm{\Omega}} \label{eq:ODE:q}, \\
    \mathbf{V}^b      &= Z \mathbf{T}^b, \\
    \bm{\Omega}^b &= \Gamma \mathbf{T}^b,
  \end{align}
\end{subequations}
where $\otimes$ denotes the quaternion product,  and $\hat{\bm{\Omega}}$ the pure quaternion formed by the vector $\bm{\Omega}$.
The transformations between the laboratory frame of reference and that of the  \ac{ABF}  are given by:
\begin{subequations} \label{eq:trans}
  \begin{align}
    \mathbf{T}^b &= R(q) \mathbf{T}, \\
    \mathbf{m} &=  R(q^\star)\mathbf{m}^b, \\
    \mathbf{V} &= R(q^\star) \mathbf{V}^b, \label{eq:ODE:trans:V} \\
    \bm{\Omega} &= R(q^\star) \bm{\Omega}^b, \label{eq:ODE:trans:Omega}
  \end{align}
\end{subequations}
where $q^\star$ is the conjugate of $q$ and $R(q)$ is the rotation matrix that corresponds to the rotation by a quaternion $q$ \cite{graf2008quaternions}.
The system of differential equations (\ref{eq:magn:torque}) and (\ref{eq:ODE:full}) is advanced in time with a fourth order Runge-Kutta integrator.

We note that when simulating multiple non-interacting \acp{ABF} in free space, we use the above \ac{ODE} system for each swimmer with the common magnetic field but different mobility coefficients and magnetic moments.

\section{Forward velocity}
\label{se:forward}

\begin{figure}
  \centering
  \begin{minipage}{0.38\columnwidth}
    \centering
    \includegraphics[width=0.99\linewidth]{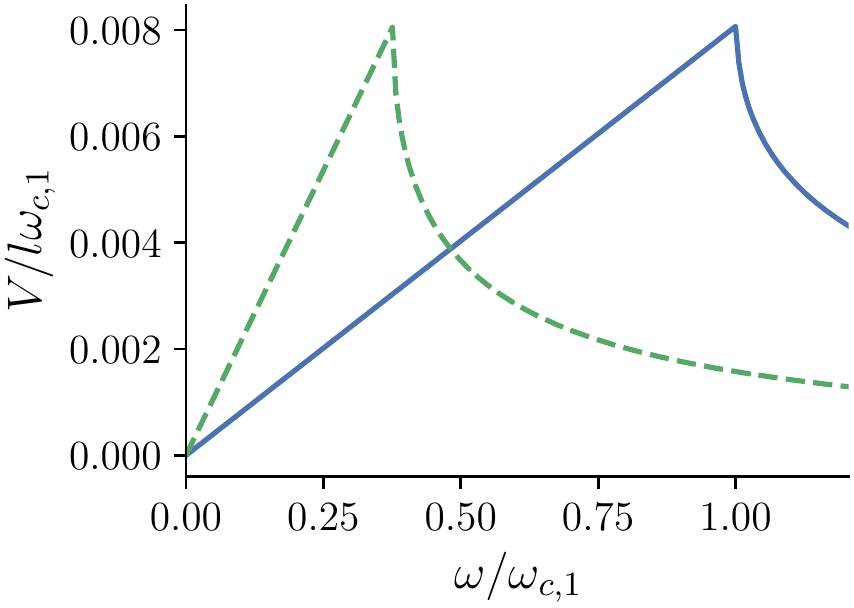}
  \end{minipage}
  \begin{minipage}[t]{0.3\columnwidth}
    \centering
    \includegraphics[width=0.99\linewidth]{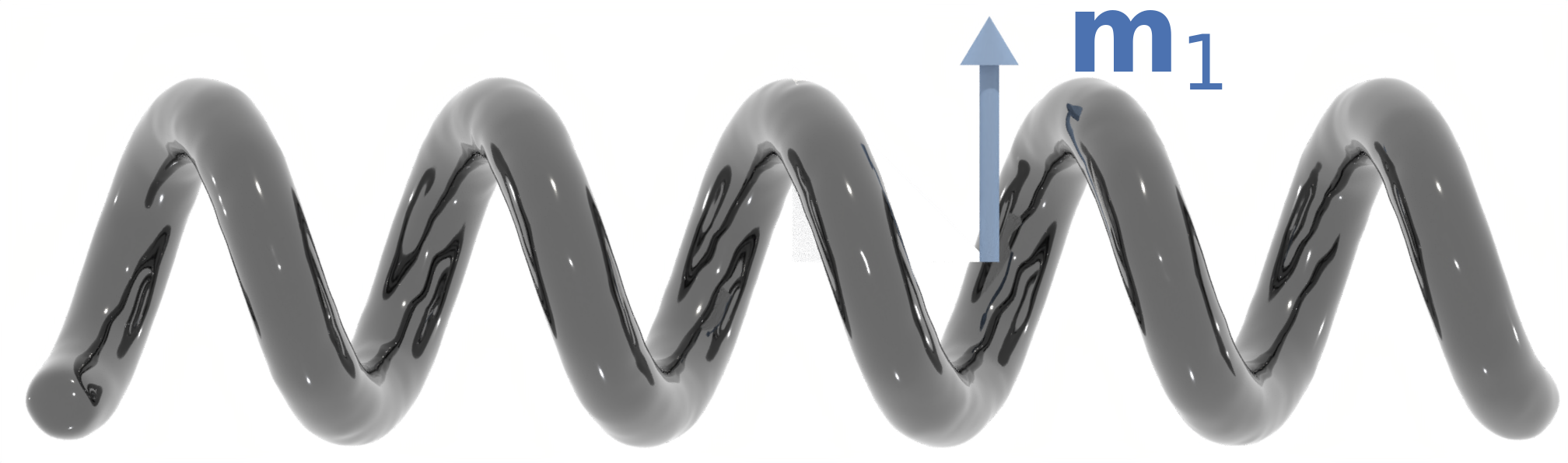} \\
    \includegraphics[width=0.99\linewidth]{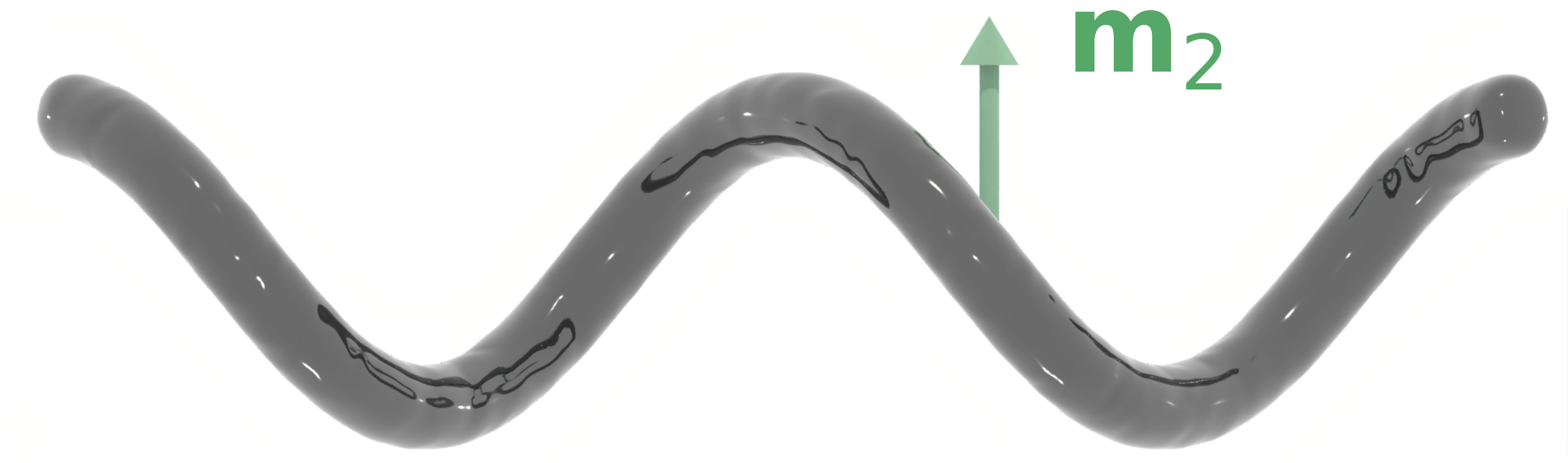}
    \vfill
  \end{minipage}

  \caption{
    Left: Dimensionless time averaged forward velocity of two \acp{ABF}, differing in shape and magnetic moment, against the field rotation frequency (in units of the step-out frequency of the first swimmer, $\omega_{c,1}$).
    Right: The \acp{ABF} geometries. The arrows represent the magnetic moment of the \acp{ABF}.
  }
  \label{fig:forward:response}
\end{figure}

\Acp{ABF} were designed to swim under a rotating, uniform magnetic field \cite{ghosh2009controlled,zhang2009artificial}.
We first study this scenario by applying the field $\mathbf{B}(t) = B \left(0, \cos{\omega t}, \sin{\omega t} \right)$ to \acp{ABF} initially aligned with the $x$ axis of the laboratory frame.
Note that in the later sections, the magnetic field is able to rotate in any direction so that the swimmers can navigate in three dimensions.
We consider two \acp{ABF} with the same length but different pitch and magnetic moments, as shown in \cref{fig:forward:response}.
In both cases, the magnetic moment is perpendicular to the helical axis of the \ac{ABF}.
Under these conditions, by symmetry of the problem, the swimmers swim along the $x$ axis.
The difference in pitch results in different coefficients of the mobility matrix and along with the different magnetic moments results in distinct propulsion velocities for the two \acp{ABF}.
For each \ac{ABF} velocity  we  distinguish a linear and a non-linear variation  with respect to the frequency of the magnetic field.
First, the \ac{ABF} rotates at the same frequency as the magnetic field and its forward velocity increases linearly with the frequency of the magnetic field, consistent with the low Reynolds approximation \cite{peyer2013bio,mhanna2014artificial,li2019helical,tierno2008}.
In the non-linear regime, the magnetic torque is no longer able to sustain the same frequency of rotation as the magnetic field. The onset of non-linearity depends on the geometry and magnetic moment of the ABF as well as the imposed magnetic field.
Indeed, the magnitude of the magnetic torque is bounded while that of the hydrodynamic torque increases linearly with the \ac{ABF} angular velocity $\Omega$.
The torque imbalance at high rotation frequencies causes the \ac{ABF} to slip, resulting in an alternating forward and backward motion (see \sm).
Increasing the frequency further increases the effective slip and accordingly  decreases the forward velocity.
The  two regimes are distinguished by the step-out frequency $\omega_c$ corresponding to the maximum forward velocity of the \ac{ABF}.

The differences in propulsion velocities for the \acp{ABF} can be exploited to control independently their trajectories.
The slope $V/\omega$ in the linear regime depends only on the shape of the \ac{ABF}.
The step-out frequency depends on both the shape and the magnetic moment (it can also be changed by varying the surface wetability of the \ac{ABF} \cite{Wang2018}).
These two properties can be chosen such that the forward velocities of two \acp{ABF} react differently to the magnetic rotation frequency (\cref{fig:forward:response}).
By changing $\omega$, it is then possible to control the relative velocities of the two \acp{ABF}: one is faster that the other in one regime while the opposite occurs in an other regime.
This simple observation constitutes the key idea for independent control of several \acp{ABF} even with a uniform magnetic field.
We remark that, while this potential has been previously identified \cite{peyer2013bio,Wang2018,bachmann2019using,khalil2018b,khalil2018}, the control of similar systems have been performed in the simple case of free space, non interacting propellers and no background flow \cite{diller2013,vach2015steering}.
To the best of our knowledge, this is the first time that such independent controlled navigation of multiple micro-swimmers is materialised in three dimensions with a complex background flow.
In the following sections we propose two methods to tackle the problem of steering \acp{ABF} towards a target in a minimal amount of time.

\section{Independent control I: semi-analytical solution}
\label{se:control}

In the absence of an external flow field, we derive a semi-analytical strategy for the  navigation of $N$ \acp{ABF} towards a particular target.
Each \ac{ABF} has a distinct magnetic moment and without any loss of generality, we set the target position of all swimmers to the origin and define the initial position of the $i^\text{th}$ \ac{ABF} as $\mathbf{x}^{(i)}$.
We assume that the time required by one \ac{ABF} to align with the rotation direction of the field is much smaller than $|\mathbf{x}^{(i)}| / v$, where $v$ is the typical forward velocity of the \ac{ABF}.
The proposed strategy consists in gathering all \acp{ABF} along one direction $\mathbf{n}_k$ at a time, such that $\mathbf{x}^{(i)} \cdot \mathbf{n}_k = 0$, $i=1,2,\dots, N$ after phase $k$.
We choose a sequence of orthogonal directions, $\mathbf{n}_k \cdot \mathbf{n}_{k'} = \delta_{kk'}$.
The choice of the orientations of $\mathbf{n}_k$ is not restricted to the basis vectors of the laboratory frame and is described at the end of this section.
In  three dimensions, the strategy consists of three phases, $k = 1,2,3$, until all \acp{ABF} have reached their target: they first gather on a plane, then on a line and finally to the target.

All \acp{ABF} are gathered along a given direction $\mathbf{n}_k$ by exploiting the different forward responses of the \acp{ABF} when we alternate the frequency of rotation of the magnetic field.
More specifically, for $N$ \acp{ABF}, the field rotates in the direction $\mathbf{n}_k$ for $t_j$ time units at frequency $\omega_{c,j}$, $j=1,2,\dots,N$, where $\omega_{c,j}$ is the step-out frequency of the $j^\text{th}$ swimmer.
We define the \textit{velocity matrix} with elements $U_{ij} = V_i\left( \omega_{c,j} \right)$, denoting the velocity of swimmer $i$ when the field rotates with the step out frequency of swimmer $j$.
We can relate the above quantities to the (signed) distances $d_j$ covered by the \acp{ABF} as
\begin{equation*}
  d_i = \sum\limits_{j=1}^N s_j t_j U_{ij},
\end{equation*}
where $s_j \in \{-1, 1\}$ determines if the field rotates clockwise/counterclockwise.
Equivalently, the vector form of the above is $\mathbf{d} = U \mathbf{\beta}$, where $\beta_{j} = t_j s_j$.
Setting $d_i = \mathbf{x}^{(i)} \cdot \mathbf{n}_k$, we can invert this linear system of equations for each phase $k$ and obtain the times spent at each step-out frequency
$\mathbf{\beta} = U^{-1} X \mathbf{n}_k$, where we have set $X_{ij} = x^{(i)}_j$.
We emphasize that this result holds only if the velocity matrix is invertible, restricting the design of the \acp{ABF} to achieve independent control.
The total time spent at phase $k$ is then given by
\begin{equation*}
  T(\mathbf{n}_k) = \sum\limits_{i=1}^N t_i = \sum\limits_{i=1}^N | \beta_i | = \norm{U^{-1} X \mathbf{n}_k}_1.
\end{equation*}
The yet unknown directions $\mathbf{n}_k$, $k=1,2,3$, are chosen to minimize the total travel time.
The directions are parameterized as $\mathbf{n}_k = R(\phi, \theta, \psi) \mathbf{e}_k$, $k=1,2,3$, where $R(\phi, \theta, \psi)$ is the rotation matrix given by the three Euler angles $\phi$, $\theta$ and $\psi$.
Note that this choice of handedness of the three directions does not influence the final result.
The optimal angles satisfy
\begin{equation*}
  \phi^\star, \theta^\star, \psi^\star = \argmin\limits_{\phi, \theta, \psi}{\sum\limits_{k=1}^3 T\left(R(\phi, \theta, \psi) \mathbf{e}_k \right) }.
\end{equation*}
We solve the above minimization problem numerically with \ac{CMA-ES} \cite{hansen2003} (see \sm for the configuration of the optimizer).

\section{Independent control II: Reinforcement Learning}
\label{se:rl}

We now employ a \ac{RL} approach to solve the problem introduced in \cref{se:control}.
Each of the $N$ \acp{ABF} is initially placed at a random position $\mathbf{x}_i \sim \mathcal{N}\left(\mathbf{x}_i^0, \sigma\right)$, $i=1,2,\dots,N$.
The \ac{RL} agent controls the magnetic field frequency of rotation and direction, and has the goal of bringing all \acp{ABF} within a small distance (here two body lengths, $d=2l$) from the target origin.
This small distance is justified by the assumption of non-interacting \acp{ABF}.
The agent sets the direction and magnitude of the magnetic field frequency every fixed time interval.
An episode is terminated if either of the two conditions occur: (a) all \acp{ABF} reached the target within a small distance $d$, or (b) the simulation time exceeds a maximum time $T_\text{max}$.
The positions $\mathbf{x}_i$ and orientations $q_i$ of the \acp{ABF} describe the state $s$ of the environment in the \ac{RL} framework.
The action performed by the agent every $\Delta t$ time encodes the magnetic field rotation frequency and orientation for the next time interval.
The reward of the system is designed so that all \acp{ABF} reach the target and the travel time is minimized.
Additionally, a shaping reward term \cite{ng1999} is added to improve the learning process.
The training is performed using VRACER, the off-policy actor critic \ac{RL} method described in \cite{novati2019a}.
More details on the method can be found in the \sm.

\section{Reaching the targets}
\label{se:results}

\begin{figure}
  \centering
  \includegraphics[width=\columnwidth]{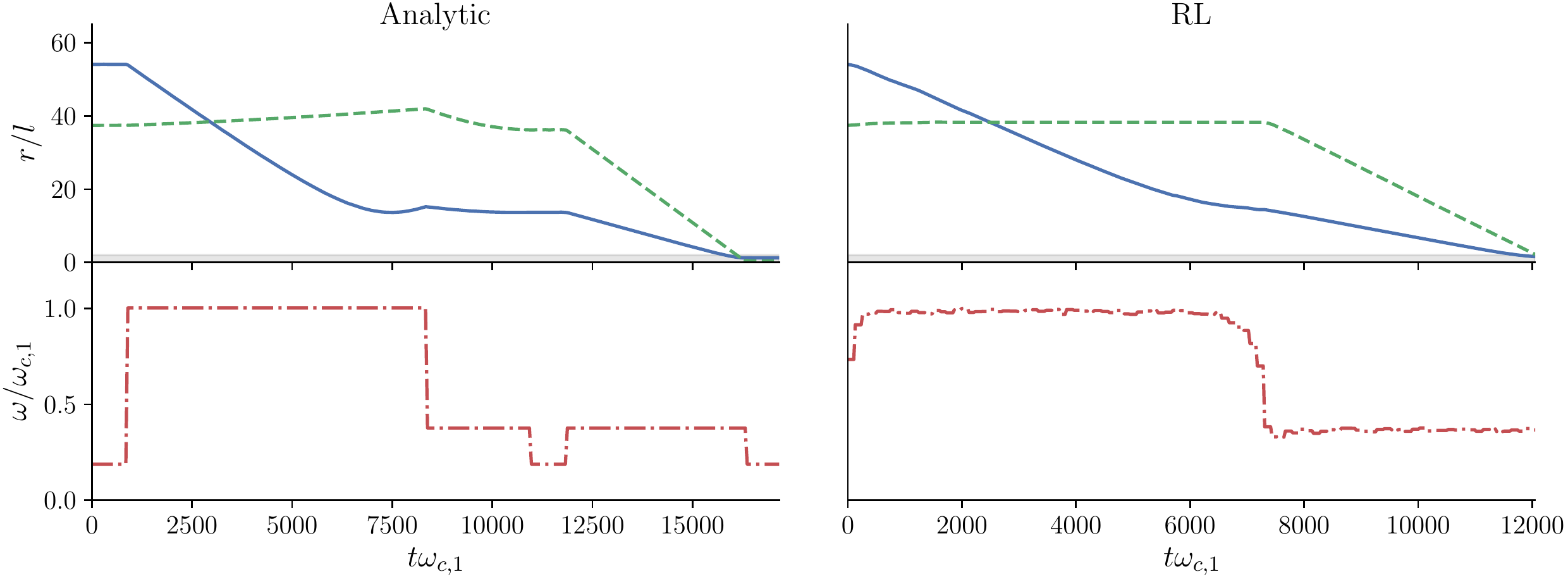}
  \caption{Distance to target of the two controlled \acp{ABF} (in units of body length $l$) against dimensionless time (\gkey{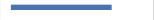} and \gkey{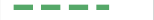}) in free space, zero background flow, and corresponding magnetic field rotation frequency (\gkey{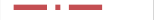}), where $\omega_{c,1}$ is the step-out frequency of the first swimmer.}
  \label{fig:freespace}
\end{figure}

\begin{figure}
  \centering
  \includegraphics[width=0.6\columnwidth]{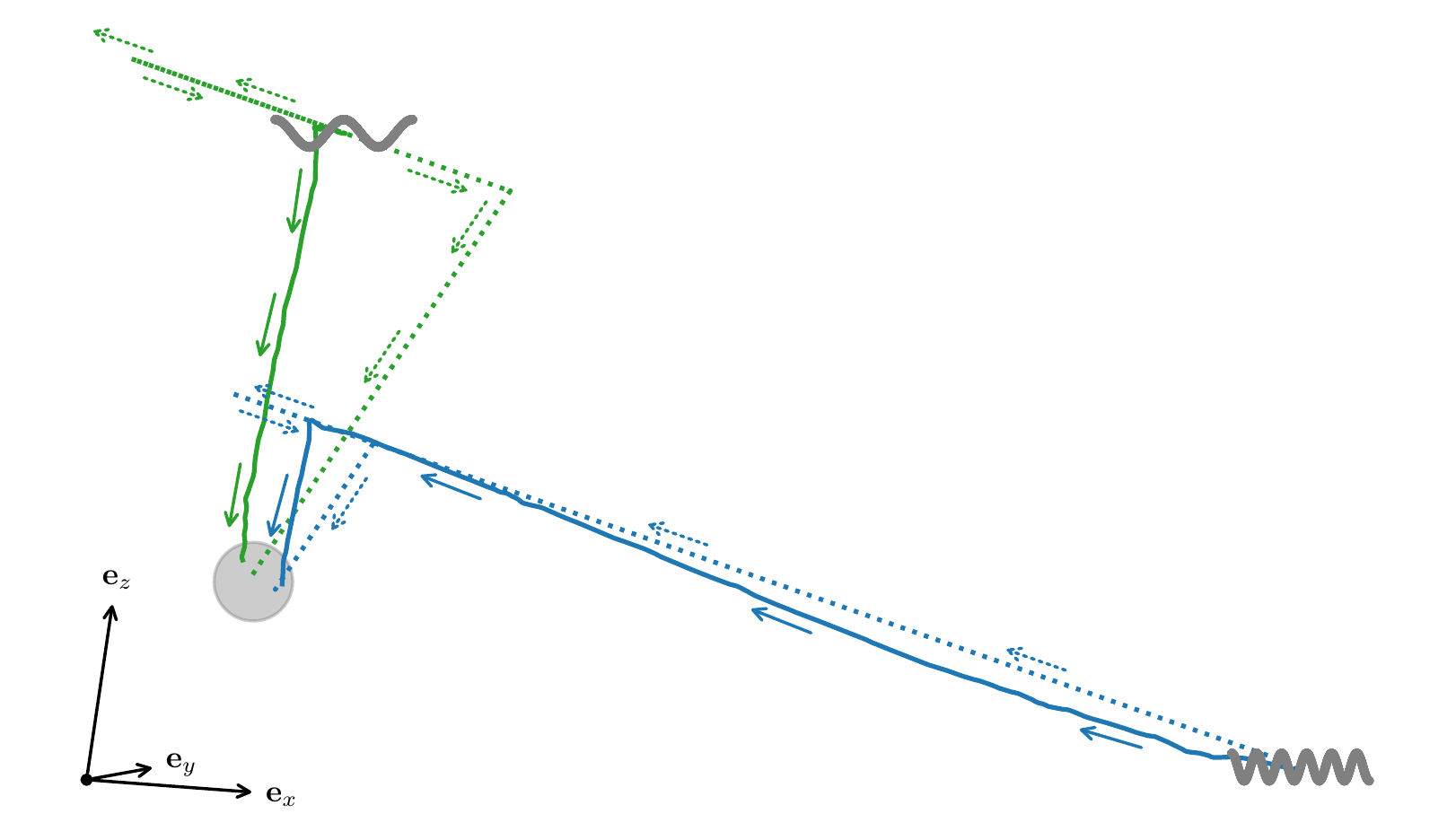}
  \caption{Trajectories of the \acp{ABF} from their initial positions (\ac{ABF} representations) to the target area (sphere) obtained with the two methods in three dimensions:
    semi-analytical (dotted lines) and \ac{RL} (solid lines).
    The arrows show the successive axes of rotation of the magnetic field.
    The size of the \acp{ABF} has been scaled up by a factor of $7$, for visualization purpose.
  }
  \label{fig:traj:noflow}
\end{figure}

In this section, we demonstrate the effectiveness of the two methods introduced in \cref{se:control,se:rl}.
We first consider 2 \acp{ABF} in free space with zero background flow.
\Cref{fig:freespace} shows the distance of the \acp{ABF} to their target over time, and the corresponding magnetic field rotation frequency for both methods.
In both cases, the \acp{ABF} successfully reach their target.
Interestingly, the rotation frequencies chosen by the \ac{RL} agent correspond to the step-out frequencies of the \acp{ABF}.
Indeed, these frequencies allow the fastest absolute velocity difference between the \acp{ABF}, so it is consistent that they are part of the fastest solution found by the \ac{RL} method.
Furthermore, the \ac{RL} trained swimmer was about $25\%$ faster than the semi-analytical swimmer.
We remark that the \ac{RL} solution amounts to first blocking the forward motion of one swimmer while the other continues swimming (see \cref{fig:traj:noflow}).
The blocked swimmer is first reoriented such that its magnetic moment is orthogonal to the plane of the magnetic field rotation, thus the resulting magnetic torque applied to this swimmer is zero.
On the other hand, the method presented in \cref{se:control} makes both \acp{ABF} swim at all time, even if one of them must go further from its target.
In such situations, the ``blocking'' method found by \ac{RL} is advantageous over the other method.

We now employ the \ac{RL} method in the case of 2 \acp{ABF} swimming in a background flow with non zero velocity.
The assumptions required for deriving the semi-analytical approach are violated and therefore we do not use this approach in this case.

In the presence of a background flow $\mathbf{u}_\infty$, \cref{eq:ODE:trans:V,eq:ODE:trans:Omega} become
\begin{align*}
  \mathbf{V} &= R(q^\star) \mathbf{V}^b + \mathbf{u}_\infty(\mathbf{x}), \\
  \bm{\Omega} &= R(q^\star) \bm{\Omega}^b + \frac{1}{2} \nabla \times \mathbf{u}_\infty(\mathbf{x}) + \frac{\lambda^2 - 1}{\lambda^2 + 1} \mathbf{p} \times \left(E(\mathbf{x}) \mathbf{p}\right),
\end{align*}
where we approximated the rotation component by the effect of the flow on an axisymmetric ellipsoid of aspect ratio $\lambda$ (Jeffery orbits).
Here $E(\mathbf{x}) = \left(\nabla \mathbf{u}_{\infty}(\mathbf{x}) + \nabla \mathbf{u}_{\infty}^T(\mathbf{x}) \right) / 2$ is the deformation rate tensor of the background flow evaluated at the swimmer's position and $\mathbf{p} = R(q^\star) \mathbf{e}_x$ is the orientation of the ellipsoid.
We used $\lambda=2$ in the subsequent simulations.
The background flow is set to the initial conditions of the Taylor-Green vortex,
\begin{equation} \label{eq:taylor:green}
  \mathbf{u}_\infty(\mathbf{r}) =
  \begin{bmatrix}
    A \cos{ax} \sin{by} \sin{cz} \\
    B \sin{ax} \cos{by} \sin{cz} \\
    C \sin{ax} \sin{by} \cos{cz}
  \end{bmatrix},
\end{equation}
with $A = B = C/2 = V_1(\omega_{c,1})$ and $a=b=-c=2\pi/50l$.
With these parameters, the maximum velocity of the background flow is larger than the maximum swimming speed of the \acp{ABF}.

\begin{figure}
  \centering
  \includegraphics[width=0.49\columnwidth]{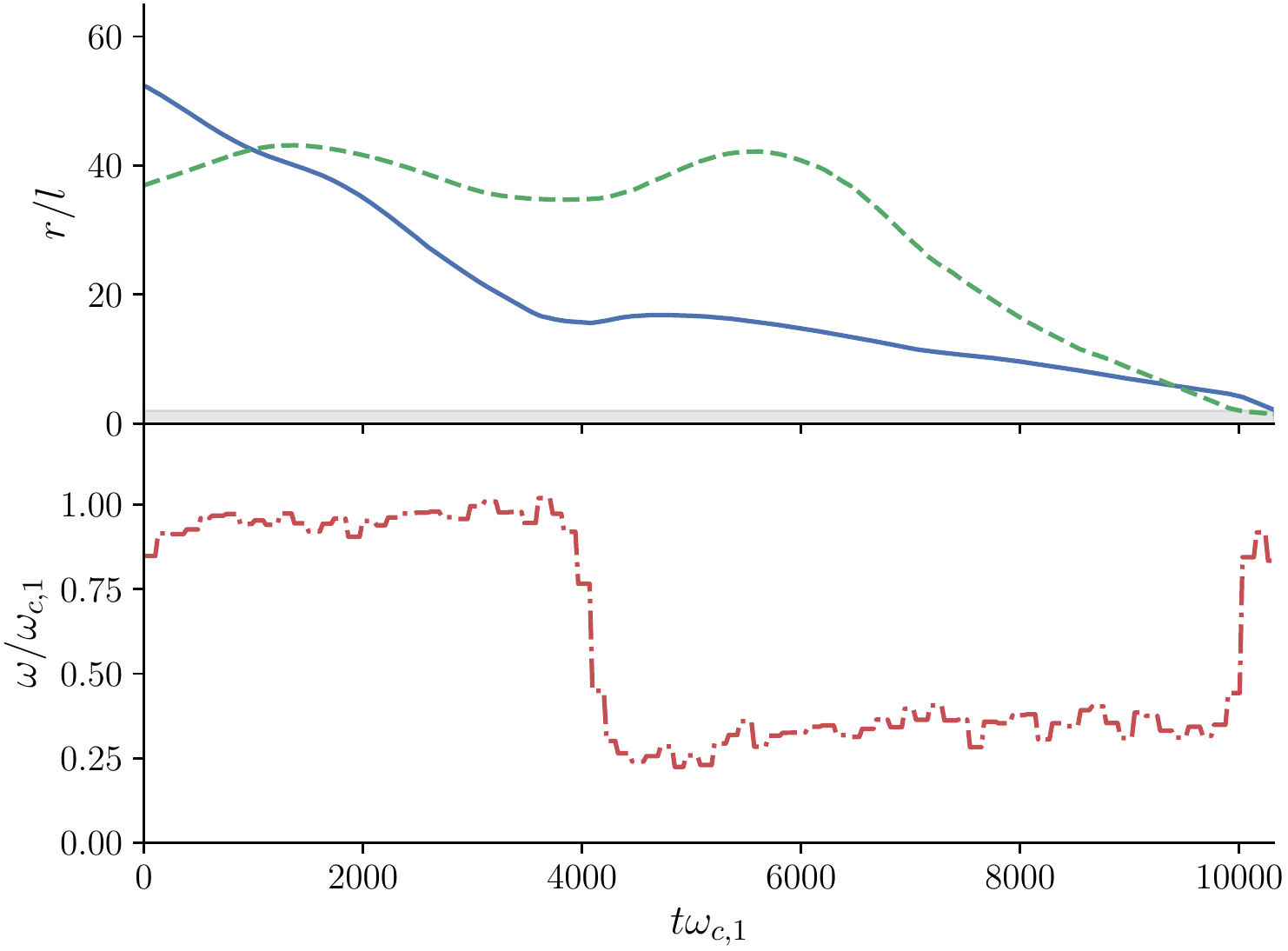}
  \includegraphics[width=0.49\columnwidth]{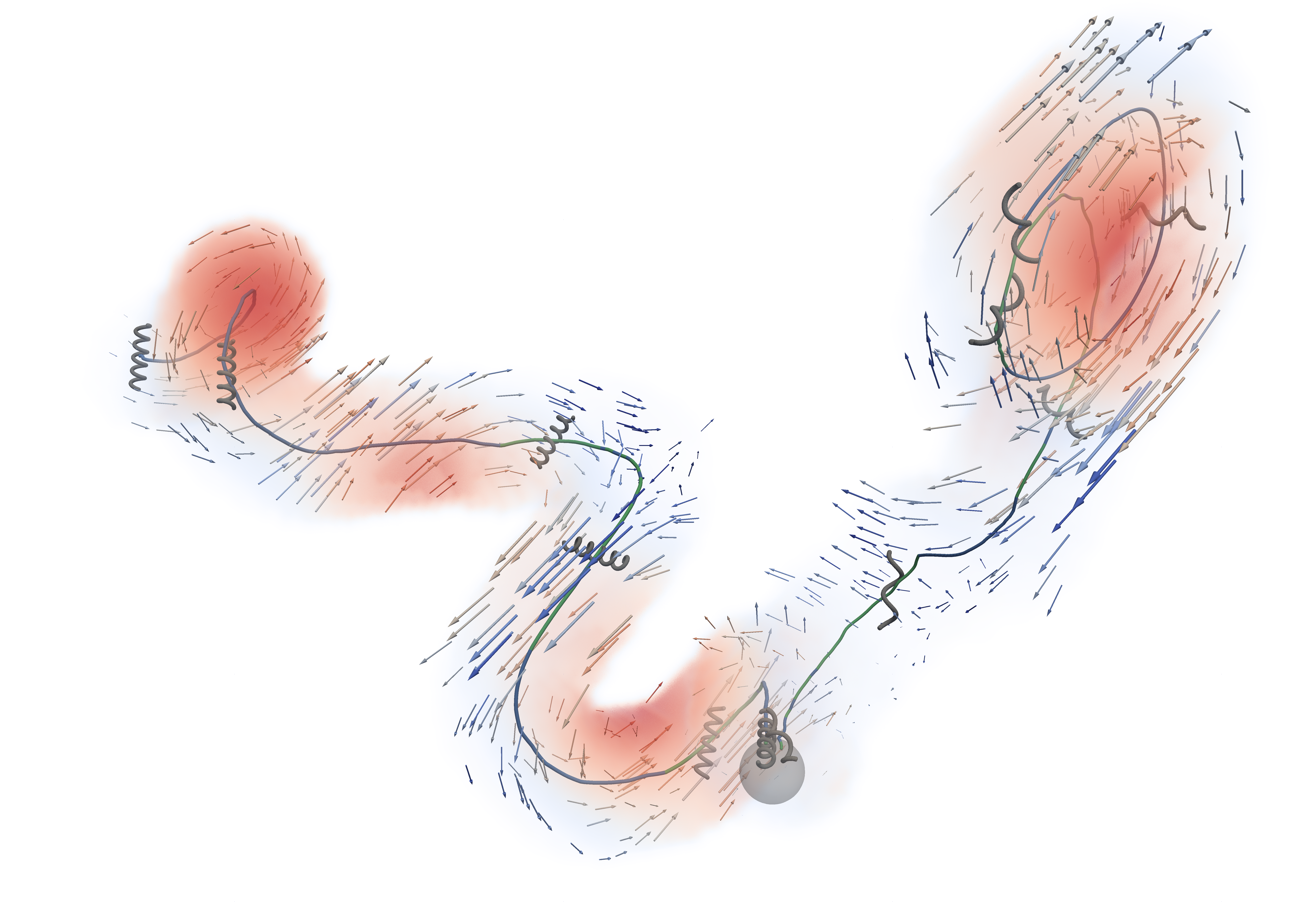}
  \caption{Left: Distance to target of the two controlled \acp{ABF} (in units of body length $l$) against dimensionless time (\gkey{01cl} and \gkey{02cl}) in free space with the background flow described by \cref{eq:taylor:green} and corresponding magnetic field rotation frequency (\gkey{03cl}), where $\omega_{c,1}$ is the step-out frequency of the first swimmer.
  Right: Trajectories of the \acp{ABF} with non-zero flow obtained with the \ac{RL} method.
    The arrows represent the velocity field and the colors represent the magnitude of the vorticity field.
    The flow field is only shown for a distance less than $4l$ from the trajectories, where $l$ is the length of the swimmers.}
  \label{fig:flow}
\end{figure}

The distances between the swimmers and the target over time are shown for the \ac{RL} method on \cref{fig:flow}.
Despite the background flow perturbation, the \ac{RL} method successfully navigates the \acp{ABF} to their target.
The magnetic action space exhibits a similar behavior as in the free space case:
the rotation frequency of the magnetic field oscillates between the step out frequencies of both swimmers and never exceeds the highest of these frequencies, where the swimming performance would degrade considerably.
The trajectories of the \acp{ABF} seem to make use of the velocity field to achieve a lower travel time: \cref{fig:flow} shows that the trajectories tend to be parallel to the velocity field.
The \ac{RL} method not only found a solution, but also made use of its environment to reduce the travel time.

\section{Robustness of the RL policy}
\label{se:robustness}

The robustness of the \ac{RL} method is tested against two external perturbations, unseen during the training phase.
In both cases, the robustness of the method is measured in terms of success rate (expected ratio between the number of successful trajectories and the number of attempts).

\begin{figure}
  \begin{center}
    \includegraphics[]{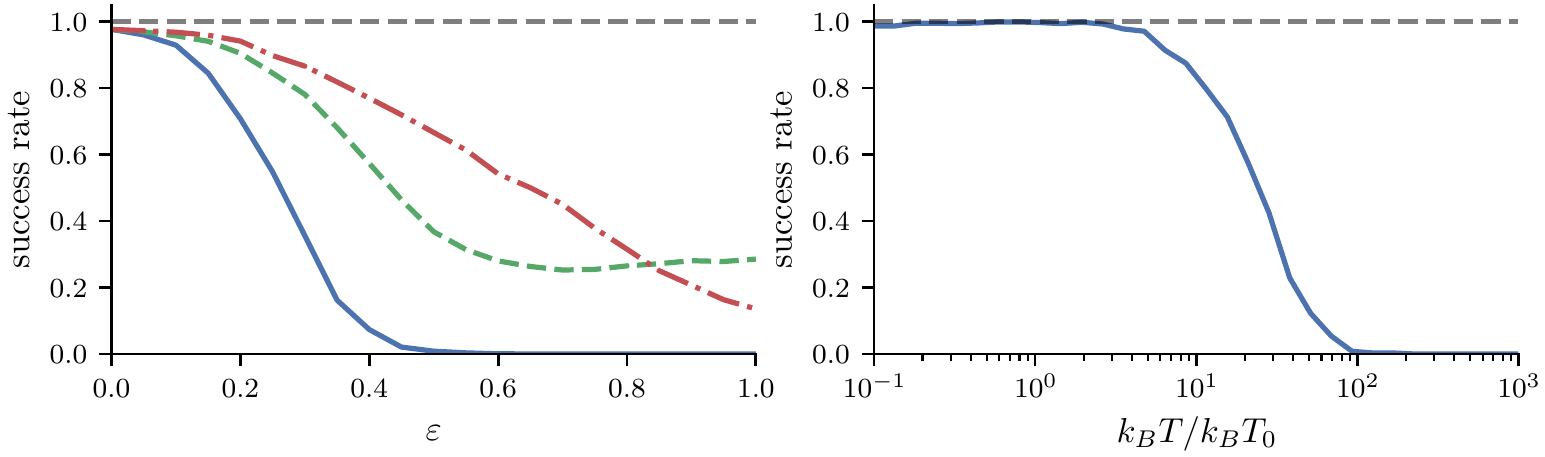}
    \caption{Left: Success rate of the \ac{RL} method to guide swimmers to their target against the flow perturbation strength $\varepsilon$, for different wave numbers,
      $p=1/2$ (\gkey{01cl}), $p=1$ (\gkey{02cl}), $p=2$ (\gkey{03cl}).
      Right: Success rate of the \ac{RL} method to steer swimmers to their target against thermal fluctuation $k_BT / k_BT_0$.
      }
    \label{fig:robustness}
  \end{center}
\end{figure}

First, a flow perturbation $\mathbf{\delta u}(\mathbf{r}) = \varepsilon \mathbf{u}_\infty(\mathbf{r} / p)$ is added to the background flow described in the previous section, where $\varepsilon$ controls the strength of the perturbation and $p$ controls the wave length of the perturbation with respect to the original one.
\Cref{fig:robustness} shows the success rate of the \ac{RL} approach against the perturbation strength $\varepsilon$ for different $p$.
For large wave lengths ($p = 2$), the \ac{RL} agent is able to successfully steer the \acp{ABF} to their target in more than $90\%$ of the cases when the perturbation strengths of less than $20\%$ of the original flow.
In contrast, the success rate degrades more sharply for smaller wave lengths ($p = 1/2$), suggesting that the method is less robust for short wave length perturbations.
The \ac{RL} policy seems more robust to perturbations with the same wavelengths as the original flow ($p=1$) for large perturbation strengths: the success rate is above $30\%$ even for large perturbations.

At small length scales, micro-swimmers are subjected to thermal fluctuations.
We investigate the robustness of the \ac{RL} policy (trained with the background flow, \cref{eq:taylor:green}) on swimmers subjected to thermal noise and background flow (\cref{eq:taylor:green}).
The thermal fluctuations are modeled as an additive stochastic term to the linear and angular velocities of each swimmer, following the Einstein relation with the mobility tensor given by \cref{eq:mobility}.
Defining the generalized undisturbed velocity $\bar{\mathcal{V}} = (\mathbf{V}, \bm{\Omega})$, the resulting stochastic generalized velocities satisfy
\begin{align*}
  \mathcal{V} &= \bar{\mathcal{V}} + \delta\mathcal{V}, \\
  \langle \delta\mathcal{V}_i \rangle &= 0, \;\; i=1,2,\dots,6, \\
  \langle \delta\mathcal{V}_i, \delta \mathcal{V}_j \rangle &= k_BT M_{ij}, \;\; i,j=1,2,\dots,6, \\
\end{align*}
where $M$ is the mobility tensor and $k_B T$ is the temperature of the system, in energy units.
The above property is achieved by adding a scaled Gaussian random noise with zero mean to the velocities at every time step of the simulation.

The success rate of the policy is shown in \cref{fig:robustness} for various temperatures $k_BT$, in units of the room temperature $k_BT_0$.
As expected, a large thermal noise causes the policy to fail at its task.
Nevertheless, this failure only occurs at relatively high temperatures: the success rate falls below $50\%$ for $k_BT > 25 k_BT_0$, which is well above the normal operating conditions of \acp{ABF}.
With temperatures below $2 k_BT_0$, the \ac{RL} method sustains a success rate above $99\%$.
We remark that this robustness is achieved successfully even with a policy trained with $k_BT = 0$.

\section{Conclusion}
\label{se:conclusion}

We have presented two methods to guide multiple \acp{ABF} individually  towards targets with a uniform magnetic field.
The semi-analytical method allows to understand the basic mechanisms that allow independent control and we
 derive the necessary condition for the independent control of multiple \acp{ABF}: their velocity matrix must be invertible, a condition that can be accommodated by suitably choosing the geometries of the swimmers.
This result may help to optimize the shapes of \acp{ABF} to further reduce the travel time.

The \ac{RL} approach can control multiple \acp{ABF} in quiescent flow as well as in the presence of a complex background flow.
Additionally, this approach is resilient to small flow perturbations and to thermal noise.
When the background flow vanishes, the \ac{RL} method recovers a very similar behavior as the first method:
the rotation frequency is alternating between the step-out frequencies of the swimmers.
Furthermore, the \ac{RL} method could reach lower travel time than the first method by blocking one swimmer while the other is swimming.
Steering an increased number of swimmers requires longer travel times according to the first method.
We thus expect that applying the \ac{RL} approach to more than two swimmers requires longer training times and might become prohibitively expensive as the number of swimmers increases.
Possible solutions to this problem might include pre-training the \ac{RL} agent with the policy found by the semi-analytical method.

The current work focused on the simplified case of non-interacting swimmers.
In practice, the \acp{ABF} may interact hydrodynamically and magnetically with each other, encounter obstacles, evolve in confined geometries or experience time varying flows.
Nevertheless, we expect the \ac{RL} method to be a good candidate to overcome these variants, in the same way as it naturally handled the addition of a background flow.


\medskip
\textbf{Acknowledgments} \par
We acknowledge insightful discussions with Guido Novati (ETHZ) and his technical support for the usage of \software{smarties}.
We acknowledge support by the European Research Council (ERC Advanced Grant 341117).

\medskip
\textbf{Conflicts of Interest} \par
The authors declare no financial or commercial conflicts of interest.

\bibliographystyle{MSP}
\bibliography{bibliography}

\end{document}


\pagestyle{fancy}
\rhead{\includegraphics[width=2.5cm]{vch-logo.png}}

\title{Supplementary Material: Independent Control and Path Planning of Microswimmers with a Uniform Magnetic Field}

\maketitle

\author{Lucas Amoudruz}
\author{Petros Koumoutsakos*}

\begin{affiliations}
L. Amoudruz, Prof. P. Koumoutsakos\\
Computational Science and Engineering Laboratory, ETH Z\"{u}rich, CH-8092, Switzerland.\\
Email: petros@seas.harvard.edu \\

L. Amoudruz, Prof. P. Koumoutsakos\\
John A. Paulson School of Engineering and Applied Sciences, Harvard University, Cambridge, MA, USA.
\end{affiliations}

\section{Neglecting the interactions between swimmers}

The \acp{ABF} experience hydrodynamic drag forces and torques.
Additionally, they are subjected to a magnetic torque from the external, imposed, magnetic field.
Furthermore, when multiple swimmers compose the system, hydrodynamic and magnetic interactions occur.
Here we estimate the magnitude of these interaction forces by considering a simplified representation of the \acp{ABF} immersed in a fluid of viscosity $\eta = \SI{e-3}{\pascal\second}$, subjected to a rotating magnetic field of magnitude $B = \SI{1}{\milli\tesla}$.
To have a rough estimate of the hydrodynamic torque, we represent the \acp{ABF} as ellipsoidal bodies of size $2a\times 2b\times 2c = \SI{10}{\micro\meter} \times \SI{2}{\micro\meter} \times \SI{2}{\micro\meter}$.
The magnetic fields rotates with angular velocity $\omega = 10 \cdot 2\pi \si{\per\second}$ which is also the step-out frequency $\omega_c$ of the \acp{ABF}.
These choices correspond to similar conditions as those found in experimental work for \acp{ABF} \cite{zhang2010artificial}.

Considering the equivalent sphere of the \ac{ABF}, with a radius $R = (abc)^{1/3}$, the hydrodynamic torque magnitude can be expressed as \cite{kim2013microhydrodynamics}
\begin{equation} \label{eq:TH}
  T_H = 8 \pi \eta R^3 \omega.
\end{equation}
At the step out frequency, the magnetic torque reaches its maximum magnitude,
\begin{equation} \label{eq:TM}
  T_M = m B,
\end{equation}
where $m$ is the magnetic moment of the sphere.
Equating \cref{eq:TH,eq:TM} gives an estimate for the magnetic moment
\begin{equation}
  m = \frac{8\pi\eta R^3 \omega}{B},
\end{equation}
hence $m \approx \SI{8e-15}{\ampere\meter^2}$.

The magnetic field generated by a dipole of moment $\mathbf{m}$ is given by
\begin{equation} \label{eq:dipole:field}
  \mathbf{B}_D(\mathbf{r}) = \frac{\mu_0}{4\pi r^3} \left( 3(\mathbf{m} \cdot \hat{\mathbf{r}}) \hat{\mathbf{r}} - \mathbf{m} \right),
\end{equation}
where $\hat{\mathbf{r}} = \mathbf{r} / r$ and $r = |\mathbf{r}|$.

Therefore, the magnetic torque exerted by one sphere of moment $\mathbf{m}_2$ to another of moment $\mathbf{m}_1$ (separated by a distance $r$) is given by
\begin{equation}
  \mathbf{T_{DD}}(\mathbf{r}) = \mathbf{m}_1\times\mathbf{B}_D(\mathbf{r})
  = \frac{\mu_0}{4\pi r^3} \left( 3(\mathbf{m}_1 \cdot \hat{\mathbf{r}}) (\mathbf{m}_2 \times \hat{\mathbf{r}}) - \mathbf{m}_1 \times \mathbf{m}_2 \right).
\end{equation}
Considering a small distance between the spheres $r = 10 a$ (five body lengths between the centers) and replacing with the numerical values, we obtain $T_{DD} / T_M \approx \num{e-5}$, hence we neglect the magnetic interactions between swimmers.
Similarly, the magnetic force caused by the gradient of the magnetic field generated by the dipole is negligible compared to that of the hydrodynamic drag.

We now estimate the hydrodynamic interactions between 2 rotating \acp{ABF} by computing the fluid velocity caused by the rotation of one at the other's location.
The magnitude of the velocity caused by the equivalent sphere of radius $R$ rotating with angular velocity $\omega$ is given by
\begin{equation}
  v_{int}(r) = \omega R^3 / r^2.
\end{equation}
This rough estimate, at a distance $r = 10 a$ (5 body lengths), gives a flow velocity of about $v_{int} \approx \SI{0.12}{\micro\meter\per\second}$.
\Acp{ABF} are able to swim at velocities of around one body length per second, that is $v \approx \SI{10}{\micro\meter\per\second}$.
Hence the velocity due to the interactions amounts to about $1.2\%$ of the swimming velocity when the swimmers are close to each other.
This velocity decreases further when the swimmers are separated by larger distances ($0.03\%$ when they are separated by 10 body lengths), which is the most common configuration in the current problem.
We therefore neglect the hydrodynamic interactions between \acp{ABF} in our simulations.

\section{Forward slip}

\begin{figure}
  \begin{center}
    \includegraphics[width=0.5\textwidth]{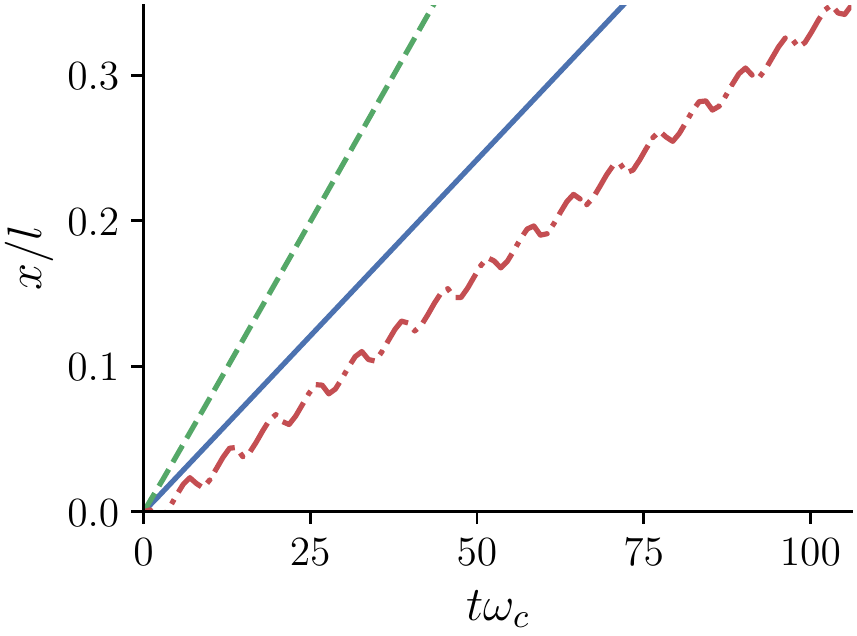}
    \caption{
      Displacement (in body lengths $l$) of an \ac{ABF} against dimensionless time ($w_c$ is the step out frequency) subjected to a constant rotating magnetic field,
      with frequency $\omega = 0.6 \omega_c$ (\gkey{01cl}), $\omega = \omega_c$ (\gkey{02cl}) and $\omega = 1.4 \omega_c$ (\gkey{03cl}).
    }
    \label{fig:forward:slip}
  \end{center}
\end{figure}

As explained in the main text, the velocity of an \ac{ABF} subjected to a uniform magnetic field rotating at constant frequency first increases linearly (its rotation follows that of the magnetic field, hence its velocity is constant).
After the frequency of rotation of the field reaches the critical value $\omega_c$, the hydrodynamic forces that would be needed to follow the magnetic field frequency are too high, causing the \ac{ABF} to ``slip''.
This can be seen on \cref{fig:forward:slip}.
Below $\omega_c$, the velocity is constant, hence the displacement of the \ac{ABF} increases linearly with time.
For higher magnetic field frequency, the \ac{ABF} slips and hence perform a back and forth motion.
The time averaged velocity hence decreases.

\section{Quaternions}
\label{se:quaternions}

Quaternions are useful mathematical tools to represent rotations in 3 dimensions \cite{graf2008quaternions}.
Unlike the rotation matrices, which have 9 components in 3 dimensions, quaternions store only 4 components, which make them also desirable on a computational point of view.
A quaternion $q$ can be represented as
\begin{equation}
  q = q_w + q_x i + q_y j + q_z k,
\end{equation}
where $q_w, q_x, q_y, q_z \in \mathbf{R}$ are real numbers and $i^2 = j^2 = k^2 = ijk = -1$.
A common notation is to  decompose the quaternion into a scalar part and a vector part,
\begin{equation}
  q = \left(q_w, \mathbf{q}\right),
\end{equation}
where $\mathbf{q} = \left(q_x, q_y, q_z\right)$.
A quaternion with zero scalar part is called a pure vector quaternion, and a quaternion with the null vector is called a pure scalar quaternion.
The conjugate of $q$ is defined as $q^\star = \left(q_w, -\mathbf{q}\right)$.
Furthermore, the norm of a quaternion is
\begin{equation}
  |q| = \sqrt{q_w^2 + q_x^2 + q_y^2 + q_z^2}.
\end{equation}

The product between two quaternions, noted $\otimes$, can be derived from the properties of $i, j, k$.
In vector notations, it is given by
\begin{equation}
  q \otimes p = \left(q_w p_w - \mathbf{q} \cdot \mathbf{p}, q_w \mathbf{p} + p_w\mathbf{q} + \mathbf{q} \times \mathbf{p}\right),
\end{equation}
where $\times$ denotes the vector product.
Rotations can be represented by unit quaternions $|q| = 1$.
The rotation of angle $\phi$ around the unit vector $\mathbf{u}$ is represented by the quaternion
\begin{equation}
  q = \left(\cos{\frac{\phi}{2}}, \sin{\frac{\phi}{2}} \mathbf{u}\right).
\end{equation}
The rotation of a vector $\mathbf{x}$ by $q$ is given in terms of quaternion product,
\begin{equation}
  y = q^\star \otimes x \otimes q,
\end{equation}
where $x = \left(0, \mathbf{x}\right)$ is the pure vector quaternion of $\mathbf{x}$ and $\mathbf{y}$ is the result of the rotation on $\mathbf{x}$.
The same operation can be written in matrix form,
\begin{equation}
  \mathbf{y} = R(q) \mathbf{x},
\end{equation}
where $R(q) = E(q)G^T(q)$ with
\begin{equation}
  E(q) =
  \begin{pmatrix}
    -q_x & q_w & -q_z & q_y \\
    -q_y & q_z & q_w & -q_x \\
    -q_z & -q_y & q_x & q_w
  \end{pmatrix},
\end{equation}
\begin{equation}
  G(q) =
  \begin{pmatrix}
    -q_x & q_w & q_z & -q_y \\
    -q_y & -q_z & q_w & q_x \\
    -q_z & q_y & -q_x & q_w
  \end{pmatrix}.
\end{equation}

\section{Calibration of the propulsion matrix from dissipative particle dynamics}
\label{se:calibration}

The mobility matrix presented in the main text links the external forces and torques applied to a rigid body with its instantaneous linear and angular velocities, relatively to the flow:
\begin{equation} \label{eq:mobility}
  \begin{bmatrix}
    \mathbf{V}^b \\
    \mathbf{\Omega}^b
  \end{bmatrix}
  =
  \begin{bmatrix}
    \Delta & Z \\
    Z^T & \Gamma
  \end{bmatrix}
  \begin{bmatrix}
    \mathbf{F}^b \\
    \mathbf{T}^b
  \end{bmatrix}.
\end{equation}
The elements of this mobility matrix depend on the shape of the body.
In this work, we estimated the mobility matrix of \acp{ABF} of different pitch but with the same length and diameter.
The body was immersed in a fluid at rest and subjected to rotations and translations with constant speed, and the hydrodynamic torques and forces were collected.
The flow solver, \texttt{Mirheo} \cite{alexeev2020a}, employs the \ac{DPD} method \cite{espanol1995statistical}, that we briefly summarize here for completeness.
The flow is discretized into $n$ particles of mass, positions and velocities $m_i$, $\mathbf{r}_i$ and $\mathbf{v}_i$, respectively, with $i=1,2\dots,n$.
We consider particles with the same mass, $m_i = m$.
They evolve in time according to Newton's law of motion,
\begin{subequations} \label{eq:dpd:motion}
\begin{align}
  \frac{d \mathbf{r}_i}{d t} &= \mathbf{v}_i, \\
  \frac{d \mathbf{v}_i}{d t} &= \frac{1}{m_i} \mathbf{f}_i,
\end{align}
\end{subequations}
where $\mathbf{f}_i$ is the total force acting on the $i^\text{th}$ particle,
\begin{align}
  \mathbf{f}_{i} &= \sum\limits_{j\neq i} \mathbf{f}^C_{ij} + \mathbf{f}^R_{ij} + \mathbf{f}^R_{ij}, \\
  \mathbf{f}^C_{ij} &= a w(r_{ij}) \mathbf{e}_{ij}, \\
  \mathbf{f}^D_{ij} &= -\gamma w_D(r_{ij}) \left(\mathbf{e}_{ij} \cdot \mathbf{v}_{ij}\right) \mathbf{e}_{ij}, \\
  \mathbf{f}^R_{ij} &= \sigma \xi_{ij} w_R(r_{ij}) \mathbf{e}_{ij},
\end{align}
where $\mathbf{r}_{ij} = \mathbf{r}_{i} - \mathbf{r}_{j}$, $r_{ij} = |\mathbf{r}_{ij}|$, $\mathbf{e}_{ij} = \mathbf{r}_{ij} / r_{ij}$ and $\mathbf{v}_{ij} = \mathbf{v}_i - \mathbf{v}_j$.
The coefficients $a$, $\gamma$ and $\sigma$ are the conservative, dissipative and random force coefficients, respectively.
$\xi_{ij}$ is a random variable with zero mean and unit variance satisfying
\begin{align*}
 & \langle \xi_{ij}(t)\rangle = 0, \\
 & \langle \xi_{ij}(t) \xi_{kl}(t')\rangle = \left(\delta_{ik}\delta_{jl} + \delta_{il} \delta_{jk}\right) \delta(t - t').
\end{align*}
The kernel $w(r) = \max{\left(1 - r / r_c, 0\right)}$ vanishes after the cutoff radius $r_c$.
The dissipative and random kernels are linked through the fluctuation-dissipation theorem
\begin{subequations} \label{eq:dpd:fluctuationdissipation}
\begin{align}
  \sigma^2 &= 2 \gamma k_BT, \\
  w_D &= w_R^2,
\end{align}
\end{subequations}
where $k_BT$ is the temperature of the fluid in energy units.
We set $w_D = w^s$ with $s=1/4$.

The \ac{ABF} is modeled as a rigid object consisting of frozen \ac{DPD} particles that move as a rigid body and interact with the surrounding solvent particles.
The frozen particles are particles chosen from a separate simulation of a \ac{DPD} solvent at rest.
Once equilibrated, the particles that are inside the volume of the \ac{ABF} are selected.
Furthermore, to enforce no-slip and no-penetrability boundary conditions, the solvent particles are bounced back from the surface of the \ac{ABF}.

The \acp{ABF} were first oriented along their principal axes and we assumed that the matrices $\Delta$, $Z$ and $\Gamma$ had non zero elements only on their diagonal.
For each \ac{ABF}, we perform 6 simulations: in each simulation, only one of the 6 components of the linear and angular velocities, $\mathbf{V}$ and $\mathbf{\Omega}$, is non zero.
The time averaged forces and torques were collected for each of these simulations, $\mathbf{F}^i$ and $\mathbf{T}^i$, where $i=1,2,\dots,6$ is the simulation index.
The coefficients of the mobility matrix are then estimated as
\begin{align}
  \Delta_{ii} &= V^i_i / F^i_i, \\
  \Gamma_{ii} &= \Omega^{i+3}_i / T^{i+3}_i,
\end{align}
for $i=1,2,3$.
The coefficient $Z_{11}$, on the other hand, is computed from the simulation of a freely translating \ac{ABF} with an imposed rotation with constant angular velocity $\Omega$ along its swimming direction in a \ac{DPD} fluid.
After reporting the averaged swimming velocity $V$, we obtain
\begin{equation}
  Z_{11} = \frac{V}{\Omega} \Gamma_{11}.
\end{equation}

All simulations are performed with a particle number density of $n_d = 10 r_c^{-3}$.
The length of the \ac{ABF} is set to $27 r_c$, resulting in more  than 3000 \ac{ABF} frozen particles in each case.
The simulation domain is periodic and has dimensions $L=270 r_c$ (ten times the length of the swimmer), which was high enough to avoid disturbances from the periodic images.

\begin{figure}
  \begin{center}
    \includegraphics[width=0.3\textwidth]{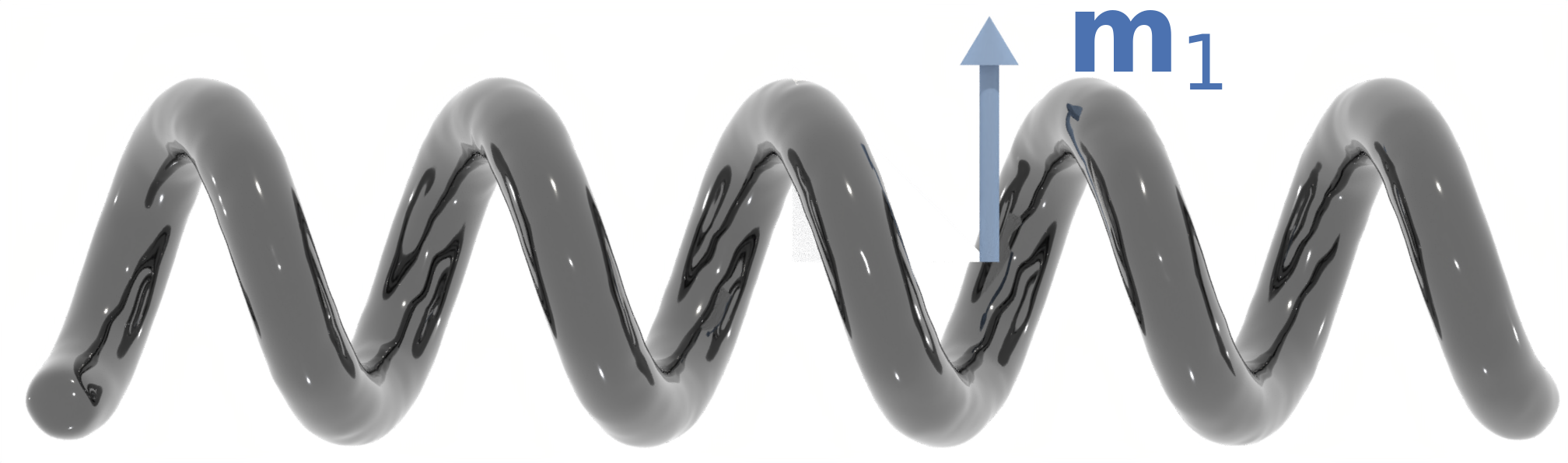}
    \includegraphics[width=0.3\textwidth]{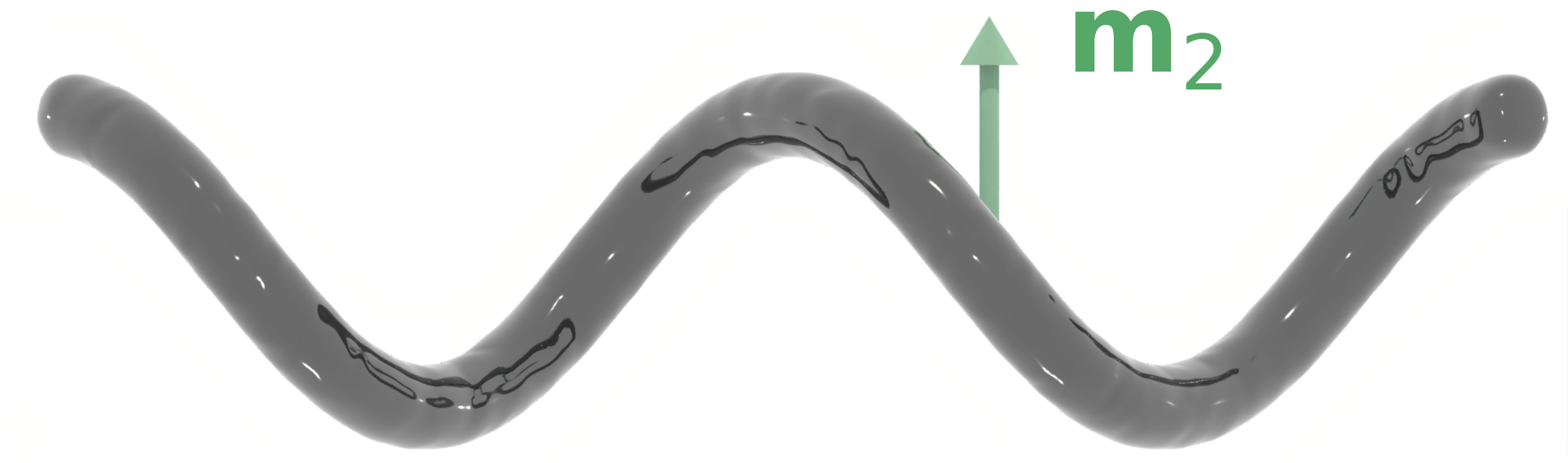}
    \caption{The geometries of the two \ac{ABF} for which the propulsion matrix has been calibrated.
      In arbitrary units, the left one has a pich of 2 and the right one has a pitch of 5.
      The arrow represents the magnetic moment $\mathbf{m}$ of the \acp{ABF}.
      \label{fig:geometries}}
  \end{center}
\end{figure}

We report the coefficients of the propulsion matrix in dimensionless form $\Delta^\star = \Delta \eta l$, $Z^\star = Z \eta l^2$ and $\Gamma^\star = \Gamma \eta l^3$, where $\eta$ is the dynamic viscosity of the fluid and $l$ is the length of the \ac{ABF}.
The \ac{DPD} simulations are run with a Reynolds number $\RE = 0.2$ and a Mach number $\MA = 0.05$ in a periodic domain of size $10l \times 10l \times 10l$ with the \acp{ABF} shown on \cref{fig:geometries}.
The nondimensional propulsion coefficients are reported in \cref{tab:propulsion}.

\begin{table}
  \begin{center}
    \begin{tabular}{cccccccc}
      Pitch & $\Delta^\star_{11}$ & $\Delta^\star_{22}$ & $\Delta^\star_{33}$ & $Z^\star_{11}$ & $\Gamma^\star_{11}$ & $\Gamma^\star_{22}$ & $\Gamma^\star_{33}$ \\ \hline
      $2$ & 0.36 & 0.32 & 0.32 & 0.096 & 11.9 & 1.7 & 1.7 \\
      $5$ & 0.41 & 0.37 & 0.37 & 0.39 & 18.2 & 2.42 & 2.42
    \end{tabular}
    \caption{Non dimensional coefficients of the propulsion matrix obtained from \ac{DPD} simulations with the two geometries shown on \cref{fig:geometries}.
      \label{tab:propulsion}}
  \end{center}
\end{table}

We choose the magnetic moment such that the maximum velocity of the swimmer equals one body length per second, which is a typical value found in experiments for a magnetic field of the order of $\SI{1}{\milli\tesla}$.
This constraint is expressed as $V_{max} = Z_{11} m B$, hence we set
\begin{equation}
  m = \frac{V_{max}}{Z_{11} B}.
\end{equation}

\section{Validation of the DPD method}

\begin{figure}[h]
  \centering
  \includegraphics[]{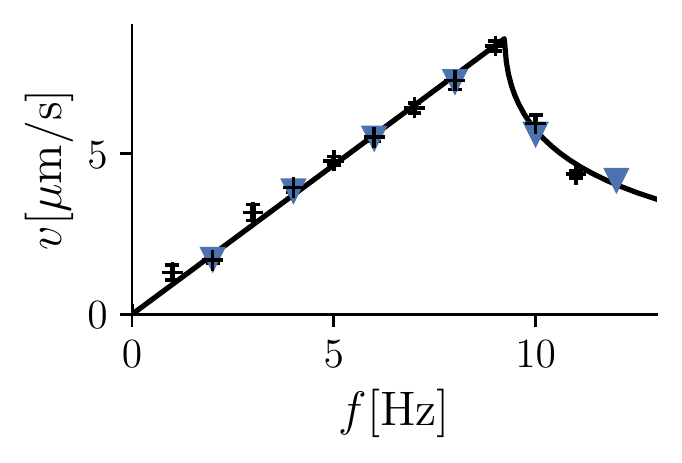}
  \caption{Swimming velocity of the swimmer against the angular frequency of the magnetic field.
    \ac{DPD} simulations (triangles) and experimental data from \cite{mhanna2014artificial} (crosses).
  }
  \label{fig:DPD:validation}
\end{figure}

The simulation of a single swimmer in 3 dimensions, in free space, is modeled similarly as in the previous section with the \ac{DPD} method.
The swimmer has a magnetic moment perpendicular to its principal component and is immersed in a uniform, rotating magnetic field $\mathbf{B}(t) = B(0, \cos \omega t, \sin \omega t)$.
The simulation employs the same resolution as described in the previous section.
The swimmer's geometry is reproduced from \cite{mhanna2014artificial} and the magnetic moment (not reported in the experiments) has been tuned to match the experimental data (m = \SI{1.0e-14}{\newton\meter\per\tesla}), with a magnetic field magnitude of $B = \SI{3}{\milli\tesla}$.
The mean swimming velocity along the $x$ axis is reported for different angular velocities $\omega$ on \cref{fig:DPD:validation} and agree well with the experimental data.

\section{Settings of the CMA-ES minimization}

The minimization of the travel time with respect to the travel directions $\mathbf{n}_k$ is performed with the gradient free method \ac{CMA-ES}.
The population size is set to $\mu=16$, the initial standard deviation is $\sigma=\pi/2$ and the minimization is stopped when the standard deviation reaches $\delta=\SI{e-5}{}$.
The first generation is centered at $(\phi, \theta, \psi) = (0, 0, 0)$.

\section{Reinforcement Learning}

The \ac{RL} method consists in an agent controlling the magnetic field through an action $\mathbf{A} = (a_\omega, a_x, a_y, a_z)$.
This action sets the frequency and direction of the rotating magnetic field for $\Delta t$ simulation time.
The angular velocity is set to $\omega = a_\omega$ and the direction is set to the unit vector  $\mathbf{u} = (a_x, a_y, a_z) / \sqrt{a_x^2 + a_y^2 + a_z^2}$.
Every $\Delta t$ time, the environment returns its state $\mathbf{s}$ and reward $r$ to the agent.
The state is composed of all positions and quaternions of the \acp{ABF}.
Due to the absence of inertia, this state describes the full system.
The reward has two components, designed to guide the \acp{ABF} towards the target in a minimal time:
\begin{equation}\label{eq:rl:reward}
  r_t = g_t - c_t.
\end{equation}
The first term is designed from reward shaping \cite{ng1999}: it increases if the \acp{ABF} come closer to the target,
\begin{equation}\label{eq:rl:reward:g}
  \begin{aligned}
    g_t &= \phi_{t-1} - \phi_{t}, \\
    \phi_t &= K_\phi \sqrt{ \sum\limits_{i=1}^N \norm{\mathbf{x}_i(t)}^2},
  \end{aligned}
\end{equation}
so that the cumulative reward (we use a discount factor $\gamma=1$), until the final simulation time $T$, depends only on the initial conditions if the \acp{ABF} reached their goal.
The constant $K_\phi$ in \cref{eq:rl:reward:g} is set such that the sum of the shaping part of the reward over one successful episodes is of the order of 1.
Therefore, this part of the cumulative reward does not affect the resulting policy \cite{ng1999}.
The second term in \cref{eq:rl:reward} is designed to penalize long travel times, $c_t = \Delta t / T_\text{max}$,
Furthermore, an additional positive term is added to the reward if all \acp{ABF} reached the target within the given distance $d_\text{success}$.
We set this constant to $K_\text{success} = 1$.
The parameters used for the two-swimmers setup are listed in \cref{tab:rl:params}.

\begin{table}[h]
  \centering
  \begin{tabular}{ll}
    \textbf{Parameter} & \textbf{Value} \\ \hline
    $\Delta t$ & $128.5 / \omega_{c,1}$ \\
    $d_\text{success}$ & $2l$ \\
    $T_\text{max}$ & $139 l/V$ \\
    $K_\text{success}$ & $1$ \\
    $K_\phi$ & $1 / 72.3 l$ \\
    $\mathbf{x}_1^0$ & $(50l, 20l, -5l)$ \\
    $\mathbf{x}_2^0$ & $(-10l, 30l, 20l)$ \\
    $\sigma$ & $3l$
  \end{tabular}
  \caption{Parameters used in the \ac{RL} setup.
    \label{tab:rl:params}}
\end{table}

We used the off-policy actor-critic algorithm \texttt{V-RACER}, implemented in the software \texttt{smarties} \cite{novati2019a,Novati2019}, with the default parameters.
A single deep neural network (with 2 hidden layers of 128 units each) maps the state space to a continuous approximation of the policy, the state value function and the action value function.
More precisely, the policy is parameterized as a gaussian whose means and variances are output of the neural network.
The neural network weights are updated through policy-gradient.
More details on the algorithm and implementation can be found in \cite{novati2019a}.

\begin{figure}
  \begin{center}
    \includegraphics[width=0.49\textwidth]{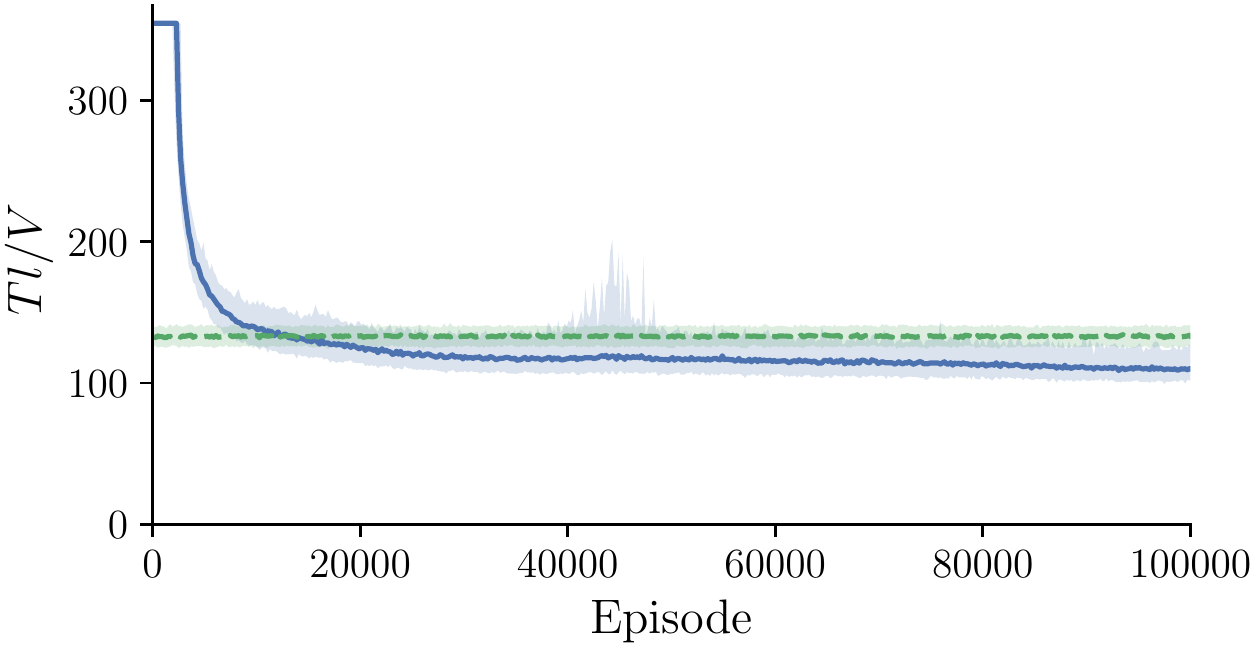}
    \includegraphics[width=0.49\textwidth]{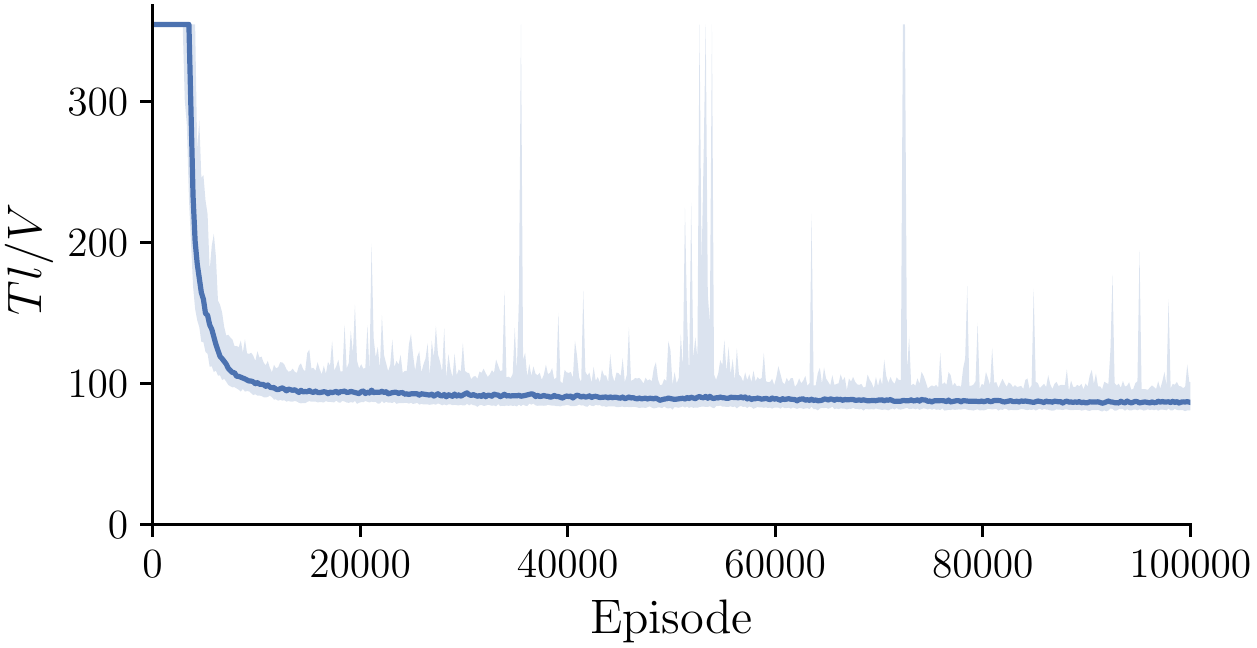}
    \caption{
      Dimensionless travel time $Tl/V$ against number of episodes during training obtained with the \ac{RL} method (\gkey{01cl}).
      The left figure corresponds to the case with 2 swimmers and no background flow.
      The time obtained with the semi-analytical method is shown for comparison (\gkey{02cl}).
      The right figure corresponds to the case with the background flow.
      $V$ is the maximum forward velocity of the \acp{ABF} and $l$ their length.
      The line corresponds to the median over 200 episodes, and the shaded area corresponds to the 0.05 and 0.95 quantiles over the same number of episodes.
    }
    \label{fig:rl:time}
  \end{center}
\end{figure}

The number of episodes needed to converge for the \ac{RL} method is of the order of $\num{2.5e4}$, as shown on \cref{fig:rl:time}.
The travel time is first equal to the maximum allowed one (when the \ac{RL} agent is not able to guide the \acp{ABF} towards the target).
After about 5000 episodes, the \acp{ABF} reach the target and the travel time decreases with the number of training episodes.
The travel time converges to a smaller value than that of the semi analytical solution.

\section{Three swimmers solution with the RL approach}

\begin{figure}
  \centering
  \includegraphics[width=\columnwidth]{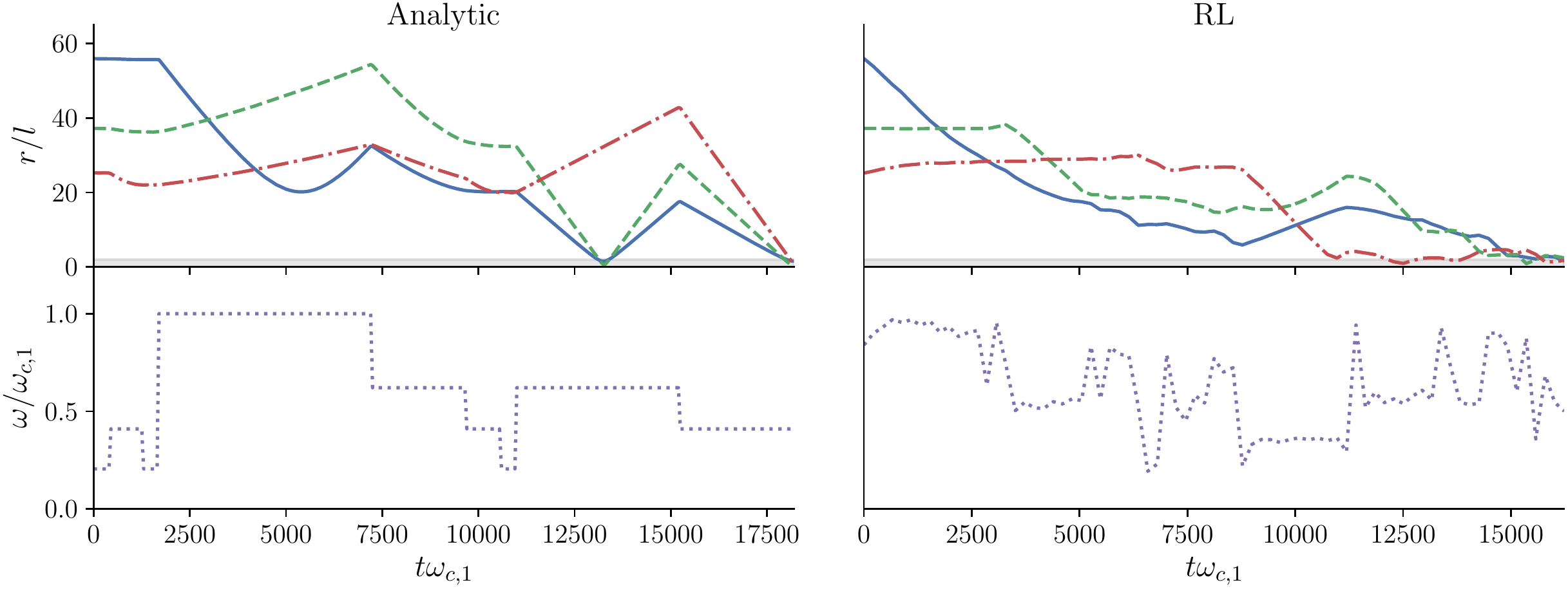}
  \caption{Distance to target of three controlled \acp{ABF} (in units of body length $l$) against dimensionless time (\gkey{01cl}, \gkey{02cl} and \gkey{03cl}) in free space, zero background flow, and corresponding magnetic field rotation frequency (\gkey{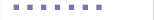}), where $\omega_{c,1}$ is the step-out frequency of the first swimmer.}
  \label{fig:freespace:3}
\end{figure}

The results obtained on two swimmers in the manuscript are here extended to three swimmers in the case of zero background flow.
\Cref{fig:freespace:3} shows the distance of the swimmers to the target.
Similarly to the two-swimmers case, the \ac{RL} method seems to alternate between the step-out frequencies of the three \acp{ABF}.

\bibliographystyle{MSP}
\bibliography{bibliography}